\documentclass[useAMS,usenatbib]{mn2e}
\usepackage{epsfig}
\usepackage{amsmath}

\newcommand{\be}{\begin{equation}}
\newcommand{\beq}{\begin{equation}}
\newcommand{\ba}{\begin{eqnarray}}
\newcommand{\ee}{\end{equation}}
\newcommand{\eeq}{\end{equation}}
\newcommand{\ea}{\end{eqnarray}}

\newcommand{\apj}{ApJ}
\newcommand{\apjl}{ApJL}
\newcommand{\mnras}{MNRAS}

\newcommand{\apjs}{ApJS}
\newcommand{\nat}{{\it Nature}}

\def\lsim{~\rlap{$<$}{\lower 1.0ex\hbox{$\sim$}}}

\def\gsim{~\rlap{$>$}{\lower 1.0ex\hbox{$\sim$}}}

\title[Dark Energy and 21cm emission]{Baryonic Acoustic Oscillations in
21cm Emission:\\ A Probe of Dark Energy out to High Redshifts}

\author[Wyithe, Loeb \& Geil]{J. Stuart B. Wyithe$^1$, Abraham Loeb$^2$  and Paul M. Geil$^1$ \\$^1$
School of Physics, University of Melbourne, Parkville, Victoria,
Australia\\$^2$ Harvard-Smithsonian Center for Astrophysics, 60 Garden St.,
Cambridge, MA 02138\\Email: swyithe@unimelb.edu.au,
loeb@cfa.harvard.edu, pgeil@physics.unimelb.edu.au}

\begin{document}


\maketitle

\label{firstpage}
\begin{abstract}

Low-frequency observatories are currently being constructed with the
goal of detecting redshifted 21cm emission from the epoch of
reionization.  These observatories will also be able to detect
intensity fluctuations in the cumulative 21cm emission after
reionization, from hydrogen in unresolved damped Ly$\alpha$ absorbers
(such as gas rich galaxies) down to a redshift $z\sim 3.5$. The
inferred power spectrum of 21cm fluctuations at all redshifts will
show acoustic oscillations, whose co-moving scale can be used as a
standard ruler to infer the evolution of the equation of state for the
dark energy.  We find that the first generation of low-frequency
experiments (such as MWA or LOFAR) will be able to constrain the
acoustic scale to within a few percent in a redshift window just prior
to the end of the reionization era, provided that foregrounds can be
removed over frequency band-passes of $\ga8$MHz. This sensitivity to
the acoustic scale is comparable to the best current measurements from
galaxy redshift surveys, but at much higher redshifts. Future
extensions of the first generation experiments (involving an order of
magnitude increase in the antennae number of the MWA) could reach
sensitivities below one percent in several redshift windows and could
be used to study the dark energy in the unexplored redshift regime of
$3.5\la z\la 12$. Moreover, new experiments with antennae designed to
operate at higher frequencies would allow precision measurements
($\la1\%$) of the acoustic peak to be made at more moderate redshifts
($1.5\la z\la3.5$), where they would be competitive with ambitious
spectroscopic galaxy surveys covering more than 1000 square
degrees. Together with other data sets, observations of 21cm
fluctuations will allow full coverage of the acoustic scale from the
present time out to $z\sim 12$.

\end{abstract}


\section{Introduction}

Measurement of the fluctuations in the intensity of redshifted 21cm emission from neutral
hydrogen promises to be a powerful probe of the reionization era
(Furlanetto et al. 2006).  The process of hydrogen reionization started
with ionized (HII) regions around the first galaxies, which later grew to
surround groups of galaxies.  Reionization completed once these HII regions
overlapped (defining the so-called {\it overlap} era) and filled-up most of
the volume between galaxies.  Detection of the redshifted 21cm signal will
not only probe the astrophysics of reionization, but also the matter power
spectrum during the epoch of reionization (McQuinn et al.~2006; Bowman et
al.~2007). 

The conventional wisdom presumes that the 21cm signal would disappear
after the {\it overlap} epoch, because there is little neutral
hydrogen left through most of intergalactic space. However, Wyithe \&
Loeb~(2007) recently demonstrated that fluctuations in the 21cm
emission would remain substantial over a range of epochs following the
end of the overlap era owing to the significant fraction by mass of
neutral hydrogen that is locked up in the dense pockets that form the
damped Ly$\alpha$ absorbers (DLAs) such as gas-rich galaxies. These
systems trace the matter power spectrum on large scales. Hence
observations of 21cm fluctuations could in principle be used as a cosmological
probe both during the reionization era and in the post reionization
IGM.

The sky temperature, which provides the limiting factor in the system
noise at the low frequencies relevant to 21cm studies, is
proportional to $(1+z)^{2.6}$, and so is a factor of $\sim 3.4
[(1+z)/5]^{2.6}$ smaller at low redshifts than for observations at
$z\sim7$.  As a result, detectability of fluctuations in 21cm emission
may not decline substantially following the overlap epoch. When
combined with the large fluctuations in redshifted 21cm emission
during the reionization era, the detectability of a 21cm power
spectrum after the end of reionization will allow the measurement of
cosmological parameters over a wide range of redshifts. Much of this
constraining power originates with redshift space distortions
(McQuinn et al.~2006), which probe cosmology through the mapping
between the vectors describing the wave number, and the emission
frequency plus the visibility. The correct mapping produces an
undistorted power spectrum. In this paper we consider baryonic
acoustic oscillations (BAO). These provide constraints on cosmology
that are related to redshift space distortions, but which are
particularly sensitive to the dark energy.

The BAO scale provides a cosmic yardstick that can be used to measure the
dependence of both the angular diameter distance and Hubble parameter on
redshift. The wavelength of the BAO is related to the size of the sound
horizon at recombination. Its value depends on the Hubble constant, and on
the matter and baryon densities. However, it does not depend on the amount
or nature of the dark energy. Thus measurements of the angular diameter
distance and Hubble parameter can in turn be used to constrain the possible
evolution of the dark energy with cosmic time. This idea was originally
proposed in relation to galaxy redshift surveys (Blake \& Glazebrook~2003;
Hu \& Haiman~2003; Seo \& Eisenstein 2003) and has since received
significant theoretical attention (e.g. Glazebrook \& Blake~2005; Seo \&
Eisenstein~2005; Seo \& Eisenstein 2007; Angulo et al. 2007). Moreover,
measurement of the BAO scale has been achieved within large surveys of
galaxies at low redshift, illustrating its potential (Cole et al.~2005;
Eisenstein et al.~2005). Galaxy redshift surveys are best suited to studies
of the dark energy at relatively late times due to the difficulty of
obtaining accurate redshifts for large numbers of high redshift
galaxies. If the dark energy behaves like a cosmological constant, then its
effect on the Hubble expansion is dominant only at $z\la1$ and becomes
negligible at $z\ga2$. In this case studies of the BAO scale at low
redshift would provide the most powerful measurement.  However, the origin
of the dark energy is not understood, and so it is not known
a~priori which redshift range should be studied in order to provide
optimal constraints on possible theories for it.

At high redshifts, baryonic acoustic oscillations in the 21cm power
spectrum should be detectable during the reionization era using future
low frequency arrays (Bowman, Morales \& Hewitt~2007). Following the
end of the reionization era, the
detection of a large number of individual galaxies in redshifted 21cm
emission could be used to trace the matter power spectrum using the
future generation of radio telescopes (Abdalla \& Rawlings~2005), in a
manner entirely analogous to optical galaxy redshift surveys. On the
other hand, 21cm observations using a low-frequency compact radio
array could also detect fluctuations in the total neutral hydrogen content
within volumes of IGM dictated by the telescope beam and frequency band
pass. Thus, in analogy to observations of the neutral IGM during the
reionization era, one could construct the power spectrum of 21cm
intensity fluctuations in the cumulative 21cm signal from all the
unresolved pockets of neutral hydrogen in the IGM, regardless of their
mass. An important advantage of studying the matter power spectrum
through detection of fluctuations in the total emission of unresolved
sources, rather than in the space density of individual sources, is
that the requirement of object detection is removed which, as we show,
allows the power spectrum to be determined at much higher redshifts
where individual sources may not be resolved in sufficient numbers.

Our goal in this paper is to investigate the feasibility of using
redshifted 21cm observations to make precise measurements of the scale
of the BAO in the matter power spectrum. We show that observations of
the acoustic scale using 21cm emission could be used to constrain the
nature of the dark energy in the unexplored redshift range of $1.5\la
z \la 6$, as well as during the reionization era, and so would be
complementary to galaxy redshift surveys.  No other probes for
precision cosmology are currently being applied to this cosmic epoch
(Corasaniti, Huterer, \& Melchiorri 2007). The outline is as
follows. In \S~\ref{21cmPS} and \ref{Uncertainty} we will describe our
approach for calculating the 21cm power spectrum, and its measurement
uncertainties.  The precision with which the scale of acoustic
oscillations can be detected using upcoming 21cm observatories will be
analyzed in \S~\ref{21cmBAO}, with the corresponding constraints on
the evolution of dark energy discussed in
\S~\ref{DarkEnergy}. Finally, we will summarize our main conclusions
in \S~\ref{conclusion}.  Unless otherwise specified we adopt the set
of cosmological parameters determined by {\it WMAP3} (Spergel et
al. 2007) for a flat $\Lambda$CDM universe.

\section{The power spectrum of 21cm fluctuations}
 \label{21cmPS}

A powerful statistical probe of the reionization era will be provided by
the power spectrum of 21cm emission which is naturally accessible to
interferometric observations such as those to be carried out by the Mileura
Widefield Array (MWA\footnote{See
http://www.haystack.mit.edu/ast/arrays/mwa/index.html}) or the Low
Frequency Array (LOFAR\footnote{See http://www.lofar.org/}). We may write
the following expression for the power spectrum of 21cm fluctuations
\begin{equation}
\label{P21}
P_{21}(k)=b_{21}(z,k)^2P(k)D(z)^2,
\end{equation}
where $P(k)$ is the primordial power spectrum of the density field as a
function of wave number $k$, extrapolated linearly to $z=0$, and $D(z)$ is
the growth factor for linear perturbations. We model the power spectrum,
including baryonic oscillations using the transfer function from Eisenstein
\& Hu~(1998). The bias like quantity $b_{21}(z,k)$, which has dimensions of
temperature, may be determined from detailed numerical simulations or from
an appropriate analytical model for bubble growth.  The bias $b_{21}$ will
in general be a function of time and scale.

We base our estimate for $b_{21}$ on the modeling presented in Wyithe \&
Loeb~(2007).  At a specified redshift, our model yields the mass-averaged
fraction of ionized regions, $Q_{\rm i}$, on various scales, $R$, as a
function of overdensity, $\delta$. We may then calculate the corresponding
21cm brightness temperature contrast
\begin{equation}
T(\delta,R) = 22\mbox{mK}\left(\frac{1+z}{7.5}\right)^{1/2}[1-Q_{\rm i}(R,\delta)]\left(1+\frac{4}{3}\delta\right),
\end{equation}
where the pre-factor of 4/3 on the overdensity refers to the spherically
averaged enhancement of the brightness temperature due to peculiar
velocities in overdense regions (Bharadwaj \& Ali~2005; Barkana \&
Loeb~2005a). Given the distribution of $\delta$ from the primordial power
spectrum of density fluctuations, we may find the probability distribution
$dP/dT$ of brightness temperature $T$ in redshifted 21cm intensity
maps. The second moment of this distribution $\langle \left(T-\langle
T\rangle\right)^{2}\rangle$ corresponds to the auto-correlation function
of brightness temperature smoothed on a scale $R$.  The effective bias can
be estimated directly from our calculation of the auto-correlation function
\begin{equation}
\label{biaseq}
b_{21}(R)^2=\frac{\langle(T-\langle T\rangle)^2\rangle}{\sigma(R)^2},
\end{equation} 
where $\sigma(R)$ is the variance of the density field smoothed on a scale
$R$.

\begin{figure}
\includegraphics[width=8.5cm]{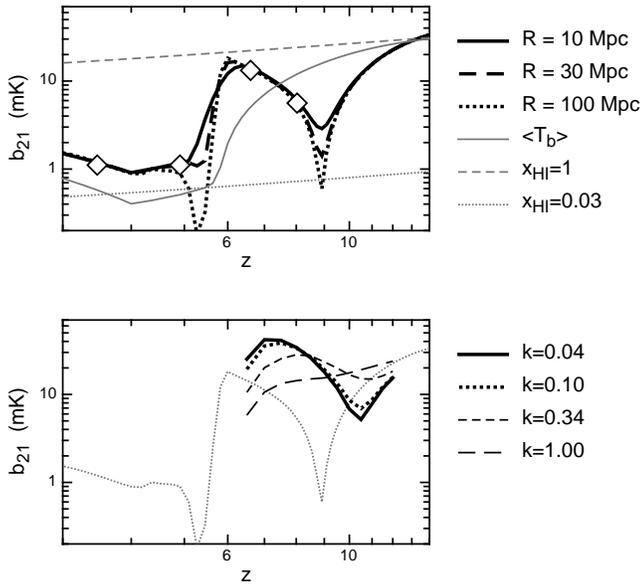} 
\caption{{\em Upper panel:} The bias ($b_{21}$) computed from our
semi-analytic model at 3 different scales (dark solid, dashed and dotted
lines). For comparison the figure also shows three curves for the bias
corresponding to the 21cm fluctuations given by spatially uniform mean
neutral fractions, with fluctuations due to the density field. In the three
cases the neutral fractions are equal to 1 (fully neutral, dashed grey
line), 0.03 (the neutral fraction in DLAs at $z\la5$, dotted grey line),
and the mean neutral fraction from our model (solid grey line).  {\em Lower
panel:} The bias ($b_{21}$) computed from our numerical model at 4
different wave numbers $k$ (dark solid, dotted, short-dashed and
long-dashed lines). The semi-analytic estimate at $R=100$Mpc is also shown
for comparison.}
\label{fig1}
\end{figure}

The upper panel of Figure~\ref{fig1} shows the bias ($b_{21}$) computed
from our semi-analytic model.  Curves are shown for model fluctuations
computed at three different scales, each of which is larger than the
typical bubble scale during the reionization era. For comparison, the
figure also shows three curves for the bias corresponding to the 21cm
fluctuations given by spatially uniform mean neutral fractions, with
fluctuations due to the density field. In the three cases, the neutral
fractions are equal to 1 (fully neutral), 0.03 (the neutral fraction in
DLAs at $z\la5$), and the mean neutral fraction from our model. We note
that our analytic estimate of the bias is a weighted average of the bias on
scales less than $R$, whereas the bias in equation~(\ref{P21}) should be
computed at a particular scale. The bias computed using our analytic model
is quite insensitive to scale for $R\ga30$Mpc, indicating that on large
scales the bias is independent of wave number $k$.

Following the end of reionization (i.e. $z\la6$) there are no longer
separate ionized bubbles as most of the IGM is ionized. Wyithe \&
Loeb~(2007) have shown that the skewness of the 21cm intensity
fluctuation distribution will be small during the post overlap epoch,
which implies a linear relation between the matter and 21cm
power spectra. Moreover, since the mean free path of ionizing photons
becomes very large in the post overlap IGM, the ionization field is
smooth. As a result the relation between the density and 21cm emission can be
reliably estimated using the semi-analytic model.

However, during the reionization epoch the relation between the power
spectrum of 21cm fluctuations and the underlying matter power spectrum
is complex, and in the late stages is dominated by the formation of
large ionized bubbles (Furlanetto et al.~2004). 
Thus prior to the end of reionization one must be careful
about applying equation~(\ref{biaseq}) to estimate the 21cm power
spectrum in equation~(\ref{P21}). On very large scales such
as those corresponding to the scale of baryonic acoustic oscillations,
the fluctuations should average over many bubbles so that the 21cm
power spectrum is again expected to be linearly related to the matter
power spectrum (McQuinn et al.~2006). Indeed, the largest discrete HII
region that could be observed during the reionization era is much
smaller than the BAO scale (Wyithe \& Loeb~2004). Redshift space
distortions can be ignored in measuring the acoustic oscillations
scale during reionization, since the characteristic peculiar velocity
is less than one percent of the Hubble velocity on this large scale.

We have computed the matter and 21cm power spectra from a hybrid
simulation (e.g. Messinger \& Furlanetto~2007) using both analytical
and numerical techniques (Geil \& Wyithe~2007). At each of redshifts
$z=12$, $z=8$ and $z=6.5$ we computed the 3-dimensional ionization
field within simulation boxes of side length 3000 co-moving Mpc. The
simulations were computed with $256^3$ resolution elements in each
case.  For these simulations we used an input power spectrum which
includes the BAO signal. At $z=12$, $z=8$ and $z=6.5$, the simulations
have global neutral fractions of 98\%, 48\% and 11\% respectively.  The
hybrid scheme of Geil \& Wyithe~(2007) is able to compute the neutral
fraction within pixels that are larger than the typical size of HII
regions. This feature allows us to compute power spectra at very large
scales even early in the reionization era when no HII regions are
resolved.

\begin{figure*}
\includegraphics[width=18cm]{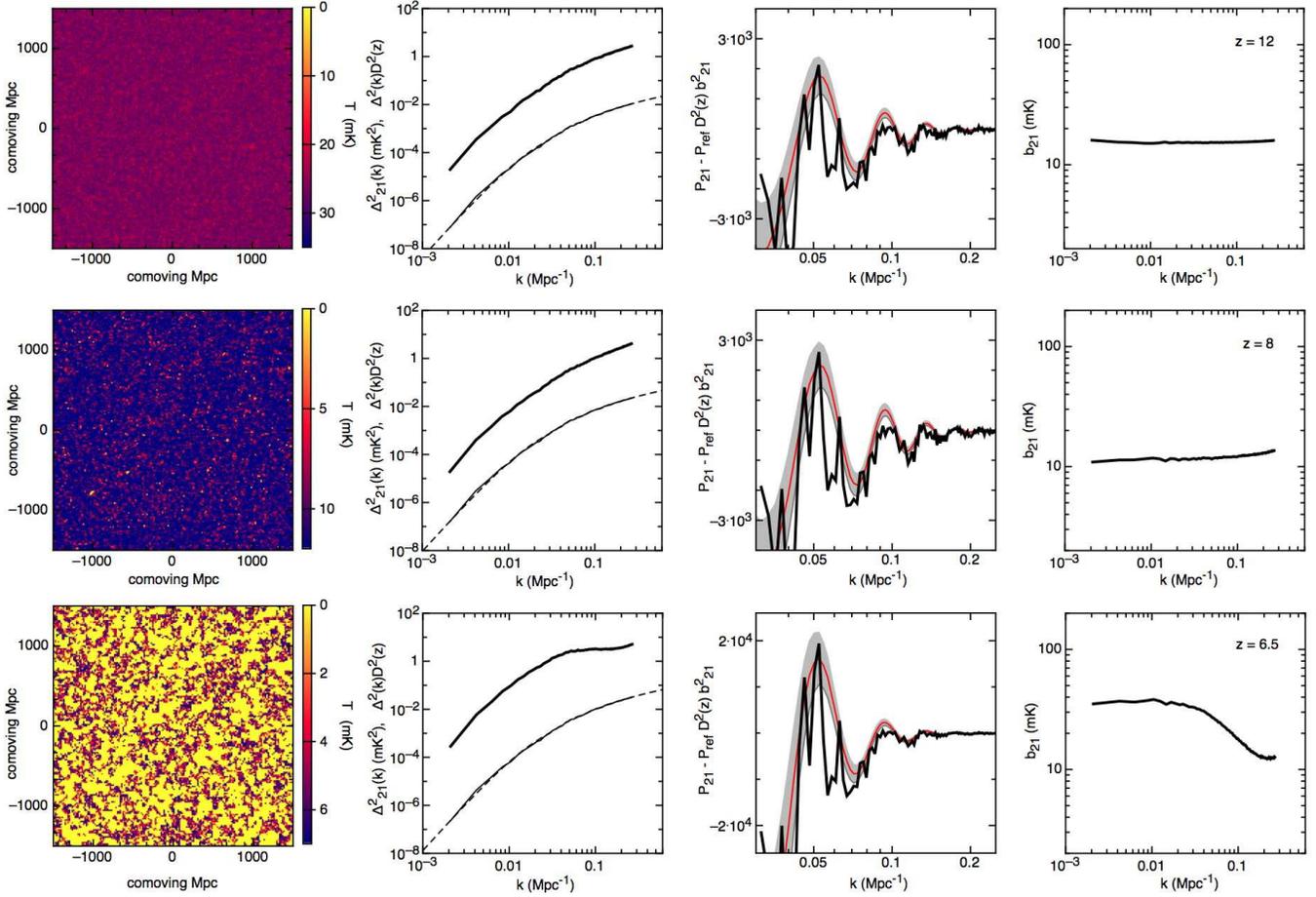} 
\caption{Examples of the 21cm power spectra during reionization. {\em Left
panels:} Maps of the 21cm emission from slices through the numerical
simulation boxes, each 3000 co-moving Mpc on a side with a thickness of 12
co-moving Mpc. In these maps yellow designates the absence of redshifted
21cm emission. {\em Central-left panels:} The corresponding matter power
spectra multiplied by the growth factor squared (thin solid lines) as well
as the 21cm (thick solid lines) power spectra computed from the simulation
box. The input co-moving power spectrum $P$ (also multiplied by the growth
factor squared) is shown for comparison (short-dashed lines). {\em
Central-right panels:} The baryonic wiggle component of the simulated 21cm
power spectrum. The curves (thick dark lines) show the difference between
the simulated 21cm power spectrum, and a reference no wiggle 21cm power
spectrum computed from the theoretical no wiggle reference matter power spectrum
multiplied by the square of the product between the bias and the growth
factor [i.e. $P_{21}-P_{\rm ref}b_{21}^2D^2$].  For comparison, the red
lines show the difference between the input matter and the no wiggle
reference matter power spectra, multiplied by the bias and growth factor
squared [i.e. $(P-P_{\rm ref})b_{21}^2D^2$]. The grey band surrounding this
curve shows the level of statistical scatter in realizations of the power
spectrum due to the finite size of the simulation volume. {\em Right
panels:} The scale dependent bias ($b_{21}$).  The upper, central and lower
panels show results at $z=12$, $z=8$ and $z=6.5$, which have global neutral
fractions of 98\%, 48\% and 11\% respectively in the model shown.  }
\label{fig2}
\end{figure*} 

The left hand panels of Figure~\ref{fig2} show the 21cm emission from
12Mpc slices through the simulation boxes at each redshift. The higher
redshift example ($z=12$) is early in the reionization era, and shows
no HII regions forming at the resolution of the simulation (i.e. the
IGM does not contain ionized bubbles with radii $\ga5$ co-moving
Mpc). The fluctuations in the 21cm emission are dominated by the
density field at this time. The central redshift ($z\sim8$) shows the
IGM midway through the reionization process, and includes a few HII
regions above the simulation resolution. The lower redshift example is
just prior to overlap, when the IGM is dominated by large percolating
HII regions, which are well resolved at the $5$Mpc resolution of our simulation.

The aim of this paper is to investigate the utility of 21cm
observations of the BAO scale as a probe of dark energy at high
redshift.  It is therefore important to demonstrate that the
percolation process does not wash out the signature of BAOs in the 21cm
power spectrum towards the end of the reionization era. In the central
left panels of Figure~\ref{fig1} we show the corresponding matter
[$\Delta^2=k^3P/(2\pi^2)$] and 21cm [$\Delta_{21}^2=k^3P_{21}/(2\pi^2)$]
power spectra. The $z=6.5$ simulation exhibits a shoulder in the 21cm
power spectrum at a scale corresponding to the characteristic bubble
size. This shoulder is also seen in analytic models of bubble growth
(Furlanetto, Zaldarriaga \& Hernquist~2004). In the two higher
redshift cases this shoulder is located at a scale below the simulation
resolution. We may use these simulations to estimate the bias
[$b_{21}(k)$] which is calculated from the square root of the ratio
between the simulated 21cm power spectrum and the simulated matter power
spectrum (using equation~\ref{biaseq}). In the right hand panels of
Figure~\ref{fig2} we show the resulting values of $b_{21}(k)$ as a
function of scale.

To illustrate the presence of BAOs in 21cm power spectra computed using
these simulations, we have also calculated the difference between the 21cm
power spectrum constructed from the simulation and a no wiggle reference
21cm power spectrum [i.e. $P_{21}(k) - P_{\rm 21,ref}(k)$]. This difference
is plotted in the central right panels of Figure~\ref{fig2}. To construct
$P_{\rm 21,ref}(k)$ we have multiplied a theoretical no wiggle reference matter power
spectrum ($P_{\rm ref}$) by the bias computed from the same simulation box
[i.e. $P_{\rm 21,ref}(k)=P_{\rm ref}(k)b_{21}(k)D(z)^2$]. Note that since
the boxes at each redshift were generated using the same realization of the
matter power spectrum, the noise is correlated between the resulting 21cm
power spectra. For comparison we also plot the difference between the
theoretical matter power spectrum and the theoretical no wiggle reference matter power
spectrum, multiplied by the bias and growth factor squared [i.e. $(P-P_{\rm
ref})b_{21}^2D^2$, which is shown by the red line in
Figure~\ref{fig2}]. The grey band around this line illustrates the level of
statistical scatter in realizations of the power spectrum due to the finite
size of the simulation volume (e.g. Peacock \& West~1992). These
simulations demonstrate that the 21cm power spectrum will exhibit BAOs
throughout the reionization epoch including the percolation phase of HII
regions.

Our simulations show that $b_{21}$ will be constant on scales much larger
than the characteristic size of HII regions, but scale dependent at larger
values of $k$. The critical $k$-scale below which $b_{21}$ is nearly
constant moves to smaller values of $k$ as reionization proceeds and the
ionized bubbles grow in size.  Early in the reionization process our
results show that there is only a very weak dependence of $b_{21}$ on scale
since the fluctuations are driven by the density field.  Indeed, our
modeling shows the scale dependence of $b_{21}$ to be weak on the scales of
interest for BAO at all times until the end of the reionization
era. At this time, just prior to the full overlap between HII
regions, the simulations show a strong dependence of $b_{21}$ on scale for
$k\ga0.05-0.1~{\rm Mpc^{-1}}$ (comparable to the BAO scale) due to the
formation of large bubbles
\footnote{Note that the bias may never be
observed to be scale dependent at values of $k$ smaller than the BAO
scale since the evolution of the power spectrum becomes more rapid
than the light crossing time of the BAO scale (Wyithe \&
Loeb~2004).}. Our simulations suggest that the scale dependent bias
distorts the observed power spectrum of BAOs (see the panels in
Figure~\ref{fig2} corresponding to $z=6.5$). However this distortion, which
causes the ratio of peak heights to increase relative to a spectrum
with constant bias, is not expected to effect the extraction of the
BAO scale. This is because in practice, the matter power spectrum
could be fitted to the data using a scale dependent bias (Seo \&
Eisenstein~2005).

The numerical values of $b_{21}$ are plotted as a function of redshift in
the lower panel of Figure~\ref{fig1}. Also shown is the semi-analytic
estimate of $b_{\rm 21}$ on a scale of $R=100$Mpc, which has been
re-plotted for comparison. Our analytic and numerical models predict
similar values of $b_{\rm 21}$ and similar behavior with redshift. In
particular, $b_{21}$ has a value of a few tens of mK both at the very
beginning and very end of reionization, despite the very different values
of the global neutral fraction. In addition, on scales much greater than
the typical bubble size there is a local minimum in the value of $b_{\rm
21}$ midway through reionization, corresponding to the shift from
fluctuations in 21cm emission being dominated by fluctuations in the
density to fluctuations in the ionization field. However, the semi-analytic
and numerical models do not agree in detail.  The semi-analytic model does
not include a Poisson component of fluctuation due to the finite number of
bubbles in a region of radius $R$ and so underestimates the value of
$b_{21}$. On the other hand, our numerical scheme does not conserve photons
[reflecting a limitation of semi-numerical models of this type (Messinger
\& Furlanetto~2007)].  As a result, while our model predicts the topology
of HII regions at a particular value of neutral fraction, it does not
correctly predict the relation between the average neutral fraction and
redshift. In the lower panel of Figure~\ref{fig1} this manifests itself as
an offset in the redshift where the local minimum of the bias is predicted
to occur.

In light of the results described above, we have chosen to use the value
for $b_{21}$ computed at $R=100$Mpc based on our semi-analytic model
through the remainder of this paper. The semi-analytic model makes a
conservatively low estimate of $b_{21}$ at all redshifts and so will yield
conservative estimates for the sensitivity of upcoming 21cm facilities to
the 21cm power spectrum.

\section{Measurement uncertainties in the power spectrum of 21cm fluctuations}
\label{Uncertainty}

Calculations of the sensitivity to the 21cm power spectrum for an
interferometer have been presented by a number of authors. We follow the
procedure outlined by McQuinn et al.~(2006), drawing on results from
Morales~(2005) and Bowman, Morales \& Hewitt~(2006) for the dependence of
the array antenna density on radius, $\rho(r)$.  The uncertainty in a
measurement of the power spectrum has two separate components. The first,
due to the thermal noise of the instrument is
\begin{equation}
\label{delN}
\Delta P_{\rm 21,N}(\vec{k}) = \left[\frac{T_{\rm sys}^2}{Bt_{\rm int}}\frac{D^2\Delta D}{n(k_\perp)}\left(\frac{\lambda^2}{A_{\rm e}}\right)^2\right]\sqrt{\frac{1}{N_{\rm c}}},
\end{equation}
where $n(k_\perp)$ is the density of baselines that observe the transverse
component of the wave-vector (Morales~2005; Bowman, Morales \&
Hewitt~2006), $T_{\rm sys}\sim 250[(1+z)/7]^{2.6}$K is the system
temperature\footnote{We assume $T_{\rm sys}$ to be dominated by the sky
throughout this paper.} of the telescope when observing the 21cm line at
redshift $z$, $B$ is the band-pass over which the measurement of the power
spectrum is made and $t_{\rm int}$ is the integration time. The quantities
$D$ and $\Delta D$ are respectively the co-moving distance to the survey
volume, and the co-moving depth of the survey volume (corresponding to the
frequency band-pass within which the power spectrum is measured). The
second component of uncertainty is due to sample variance within the finite
volume of the survey, and equals
\begin{equation}
\label{delSV}
\Delta P_{\rm 21,SV}(\vec{k}) = P_{\rm 21}(\vec{k})\sqrt{\frac{1}{N_{\rm
c}}}.
\end{equation}
The noise is evaluated within a $k$-space volume element $d^3k$. The total
noise within a finite $k$-space bin may then be obtained by integration
over the volume within the bin.  In both equations~(\ref{delN}) and
(\ref{delSV}) the quantity $N_{\rm c} = 2\pi k^2\sin \theta
dkd\theta[\mathcal{V}/(2\pi)^3]$ denotes the number of modes observed
within a $k$-space volume element $d^3k=2\pi
k^2\sin(\theta)dkd\theta$. Note that in computing $N_{\rm c}$ we have
assumed symmetry about the polar angle and expressed the wave vector
$\vec{k}$ in components of its modulus $k$ and angle $\theta$ relative to
the line-of-sight. Because wave numbers can only be observed if their
line-of-sight component fits within the observer's band-pass, we set
$N_{\rm c}=0$ if $2\pi/k\, \cos(\theta)>\Delta D$.  The number of modes
observed depends on the volume of the survey, $\mathcal{V} = D^2\Delta
D(\lambda^2/A_{\rm tile})$, where $A_{\rm tile}$ is the total physical
surface area of an antenna (this point is discussed further below).

The sensitivity to the 21cm power spectrum is dependent on both the
sensitivity of the telescope to a particular mode, and to the number of
such modes in the survey. The former is set by the effective collecting
area ($A_{\rm e}$) of each antenna element (as well as the total number of
antennae), while the latter is sensitive to the total physical area covered
by each antenna (which we refer to as $A_{\rm tile}$). For a traditional
interferometer consisting of a number of dishes in a phased array, these
two areas are approximately equivalent $A_{\rm e}\sim A_{\rm tile}$, since
the solid angle of the primary beam and the sensitivity are both
proportional to the physical collecting area of the dish. However in
constructing the above formalism for the sensitivity to the power spectrum,
we have explicitly allowed $A_{\rm e}\neq A_{\rm tile}$.  This is because
future interferometers being built to measure fluctuations from the epoch
of reionization (like the MWA) will not be comprised of dishes, but rather
a large number of 'tiles', each consisting of a phased array of $N_{\rm
dip}$ dipoles distributed over an area $A_{\rm tile}$. Since the size of
the dipole will be much lower than $\lambda$ for observations of the 21cm
line at $z\ga3.5$, the effective collecting area of each tile in this
regime is $A_{\rm e}\sim N_{\rm dip}\lambda^2/4$ (Bowman et al.~2005).
Each tile forms an electronically steerable primary beam, with solid angle
$\Omega_{\rm beam}\sim \lambda^2/A_{\rm tile}$. The MWA is designed to
observe the 21cm line from the epoch of reionization, and so has ($A_{\rm
e}\sim A_{\rm tile}$) when observing at $z\sim8$, but $A_{\rm e}<A_{\rm
tile}$ at lower redshifts\footnote{The relation between $A_{\rm e}$ and
$A_{\rm tile}$ is often expressed in terms an aperture efficiency
$\epsilon=A_{\rm e}/A_{\rm tile}$ (e.g. Morales~2005).}. In terms of
measuring a power spectrum this reduces the efficiency of the MWA at higher
frequencies. In this paper we consider the power spectrum at $1.5\la z\la6$
as well as during the epoch of reionization at $z\ga6$. When showing
results at $z>3.5$ we use the specifications of the MWA since this
observatory is already under construction and its design is not
flexible. Observations of neutral hydrogen at $z<3.5$ are not accessible to
the MWA, and so a new telescope would need to be constructed for probing
this epoch. Thus, at $z<3.5$ we assume $A_{\rm e}\sim A_{\rm tile}$,
corresponding to a telescope with an optimal design (i.e. with dipoles
spaced by $\lambda/2$) for observations at these lower redshifts.

In the case of a spherically averaged power spectrum, $P_{21}(k)$, each of
the above noise components can be computed within $k$-space volumes of
$d^3k = 4\pi k^2\Delta k$, where $\Delta k$ is a finite bin of values in $k$. 
However, the power spectrum, $P_{21}(k_{\perp},k_{\parallel})$, can also be
expressed in terms of the wave vector components that are parallel
($k_{\parallel}$) and perpendicular ($k_{\perp}$) to the line-of-sight, in
which case the components $\Delta P_{\rm 21,N}$ and $\Delta P_{\rm 21,SV}$
can be computed within $k$-space volumes of $d^3k = 2\pi k_{\perp}\Delta
k_{\perp}\Delta k_{\parallel}$.

The contamination by foregrounds provides an additional source of
uncertainty in the estimate of the power spectrum. McQuinn et al.~(2006)
have shown that it should be possible to remove the power due to
foregrounds to a level below the noise in the cosmological signal, provided
that the region of frequency band-pass from which the power spectrum is
estimated ($B$) is substantially smaller than the total band-pass available
($B_{\rm total}$). Following the approximation suggested in McQuinn et
al.~(2006), we combine the above components to yield the uncertainty in the
estimate of the power spectrum. For the spherically averaged power spectrum
we assume
\begin{equation}
\label{delsym}
\Delta P_{21}(k) = \left\{ 
\begin{array}{ll}
\frac{\Delta P_{\rm 21,SV}(k) + \Delta P_{\rm 21,N}(k)}{\sqrt{N_{\rm fields}\,B_{\rm tot}/B}}, & \mbox{if $k>k_{\rm min}$}\\ 
\infty, & \mbox{otherwise}
\end{array}
\right.
\end{equation}
while for noise in the power spectrum at $\vec{k}=(k_{\parallel},k_{\perp})$ we take
\begin{equation}
\label{delasym}
\Delta P_{21}(\vec{k}) = \left\{ 
\begin{array}{ll}
\frac{\Delta P_{\rm 21,SV}(\vec{k}) + \Delta P_{\rm 21,N}(\vec{k})}{\sqrt{N_{\rm fields}\,B_{\rm tot}/B}}, & \mbox{if $k_{\parallel}>k_{\rm min}$}\\ 
\infty, & \mbox{otherwise}
\end{array}
\right.
\end{equation}
where $k_{\rm min}=2\pi/\Delta D$. In each of equations~(\ref{delsym}) and
(\ref{delasym}) the denominator represents the number of independent
measurements made of the power spectrum within a band width $B$. The factor
$N_{\rm fields}$ is included because one independent measurement would be
made per field imaged for time $t_{\rm int}$. As part of their analysis of
foreground removal, McQuinn et al.~(2006) have found that foreground power
could be removed within a region $B$ of the observed band-pass provided $B$
is significantly smaller than $B_{\rm tot}$. However in addition McQuinn et
al.~(2006) have also found that foreground removal is not sensitive to the
location of $B$ within the total processed band pass. As a result, on
scales where the power spectrum can be measured (i.e. at $k>k_{\rm min}$)
it can be determined within $B_{\rm total}/B$ independent regions of the
total processed band pass $B_{\rm total}$.  Hence while foreground removal
will limit the scale of fluctuations that can be observed, foregrounds
should not affect the total sensitivity of the array to smaller scale
modes. We note that, as with all studies of 21cm fluctuations, the level to
which it will be possible to remove foregrounds provides the largest
uncertainty in our analysis.

\section{21cm observations of baryonic acoustic oscillations}
\label{21cmBAO}

\begin{figure*}
\includegraphics[width=15cm]{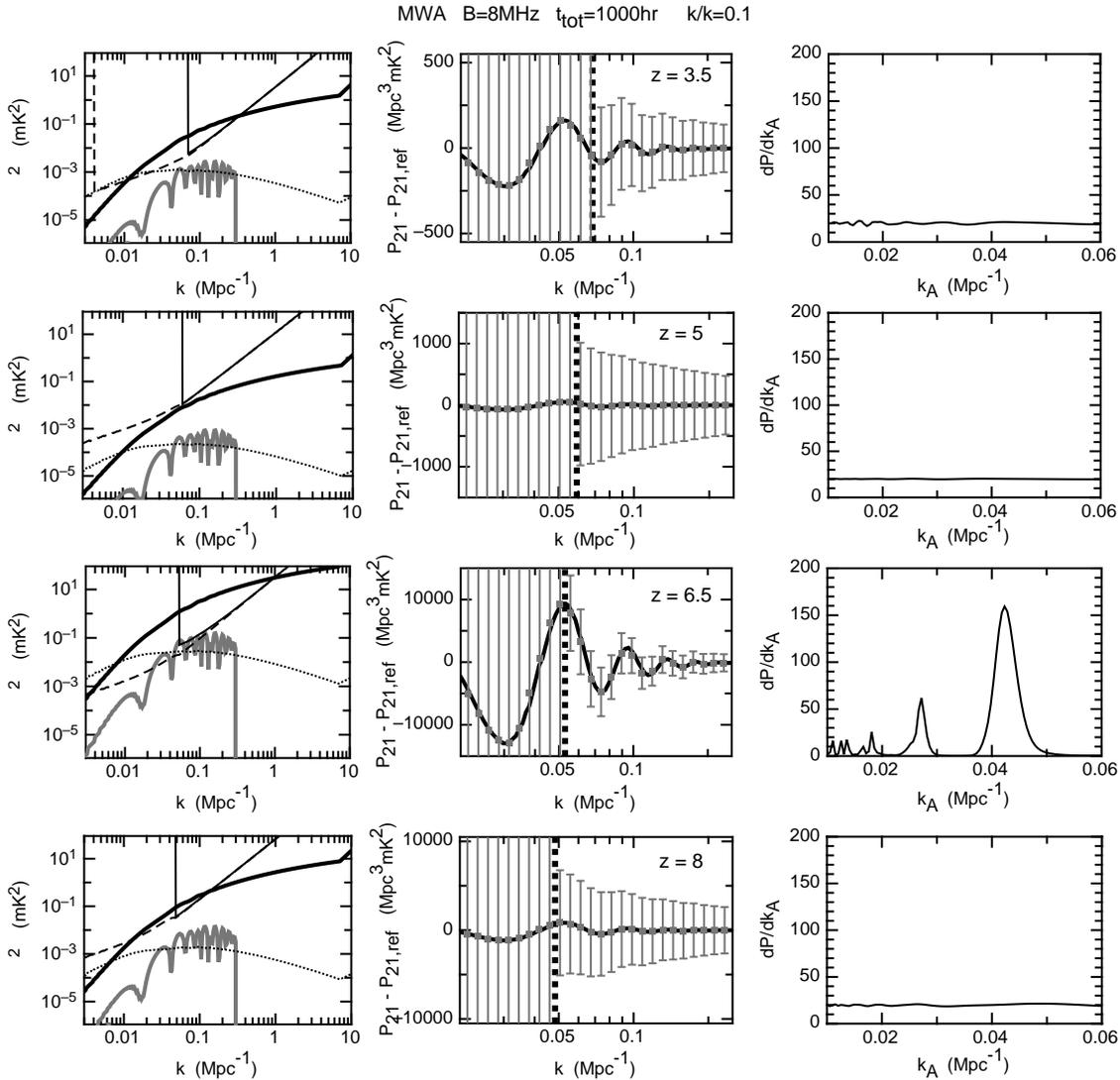} 
\caption{The power spectrum of 21cm fluctuations and measurement of the
baryonic acoustic oscillations. Results are shown at four redshifts,
$z=3.5$, 5, 6.5 and 8, and assume a low-frequency array with the
specifications of the MWA, 1000hr of integration on a single field and
foreground removal within $B=8$MHz of band-pass. {\em Left panels:} The
power spectrum $\Delta^2(k)\equiv k^3P_{21}(k)/(2\pi^2)$ of 21cm
fluctuations (thick solid lines). The absolute value of the component of
the 21cm power spectrum due to the baryonic acoustic oscillations is also
plotted at $k<0.3$Mpc$^{-1}$ (heavy grey lines). Also shown for comparison
are estimates of the noise. In each case we plot the sample variance
(dotted lines) and thermal noise (dashed lines) components of the
uncertainty within $k$-space bins of size $\Delta k=k/10$. The combined
uncertainty including the minimum $k$ cutoff due to foreground subtraction,
is also shown as the thin solid lines.  {\em Central panels:} The power
spectra with the representative smooth power spectrum subtracted. The
points with error bars show the accuracy attainable within a bin of width
$\Delta k/k=0.1$. The vertical dotted line is the wave number corresponding
to the band-pass, below which the error bars are very large.  Note that the
vertical scale is different at each redshift and has been chosen to best
illustrate the magnitude of the uncertainty relative to the amplitude of
the BAOs. {\em Right panels:} The probability distributions for the
recovered acoustic wave number $k_{\rm A}$ at each redshift. }
\label{fig3}
\end{figure*} 

\begin{figure*}
\includegraphics[width=15cm]{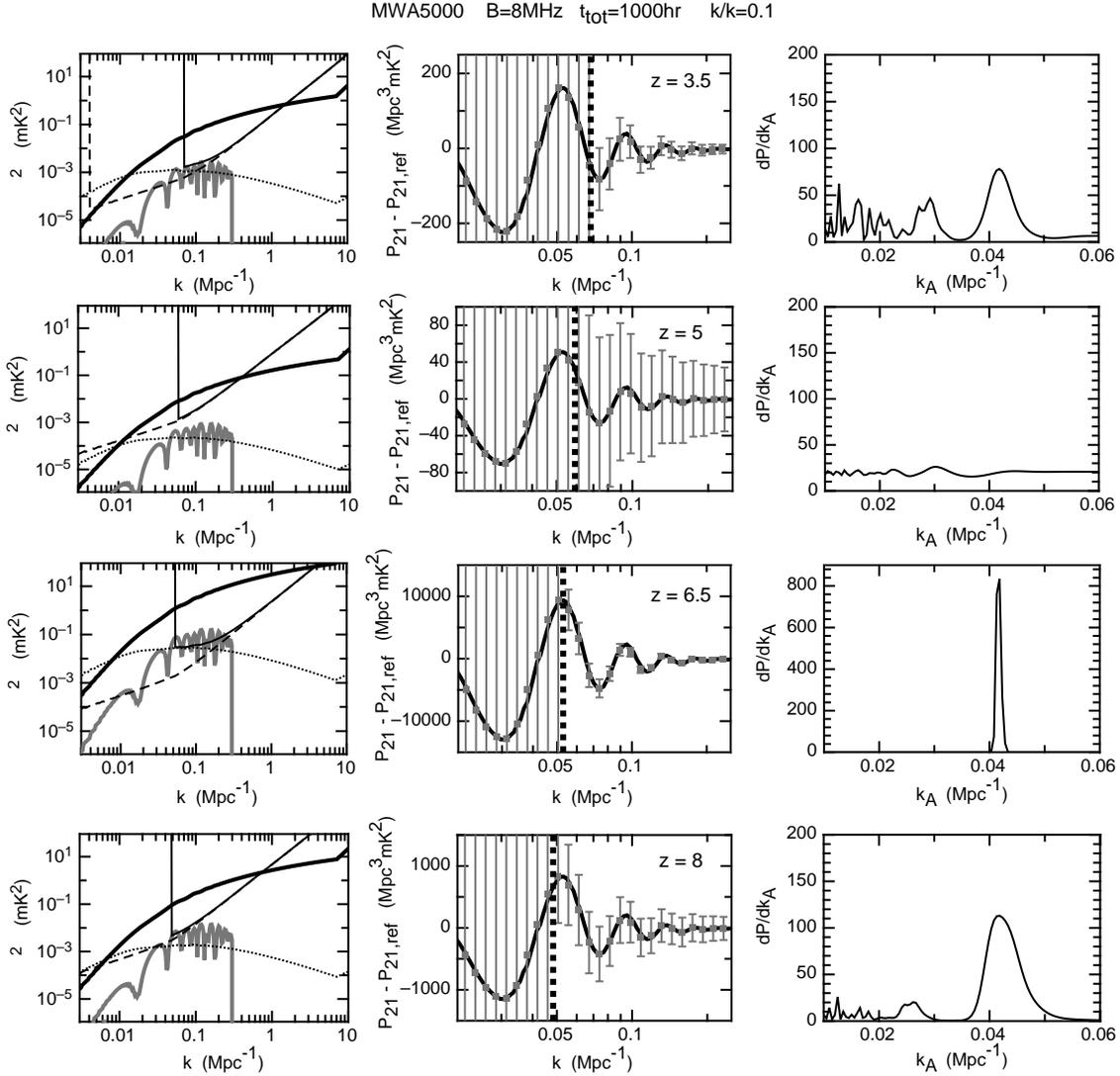} 
\caption{As in Figure~\ref{fig3}, but assuming the layout of the MWA5000. }
\label{fig4}
\end{figure*} 

\begin{figure*}
\includegraphics[width=14cm]{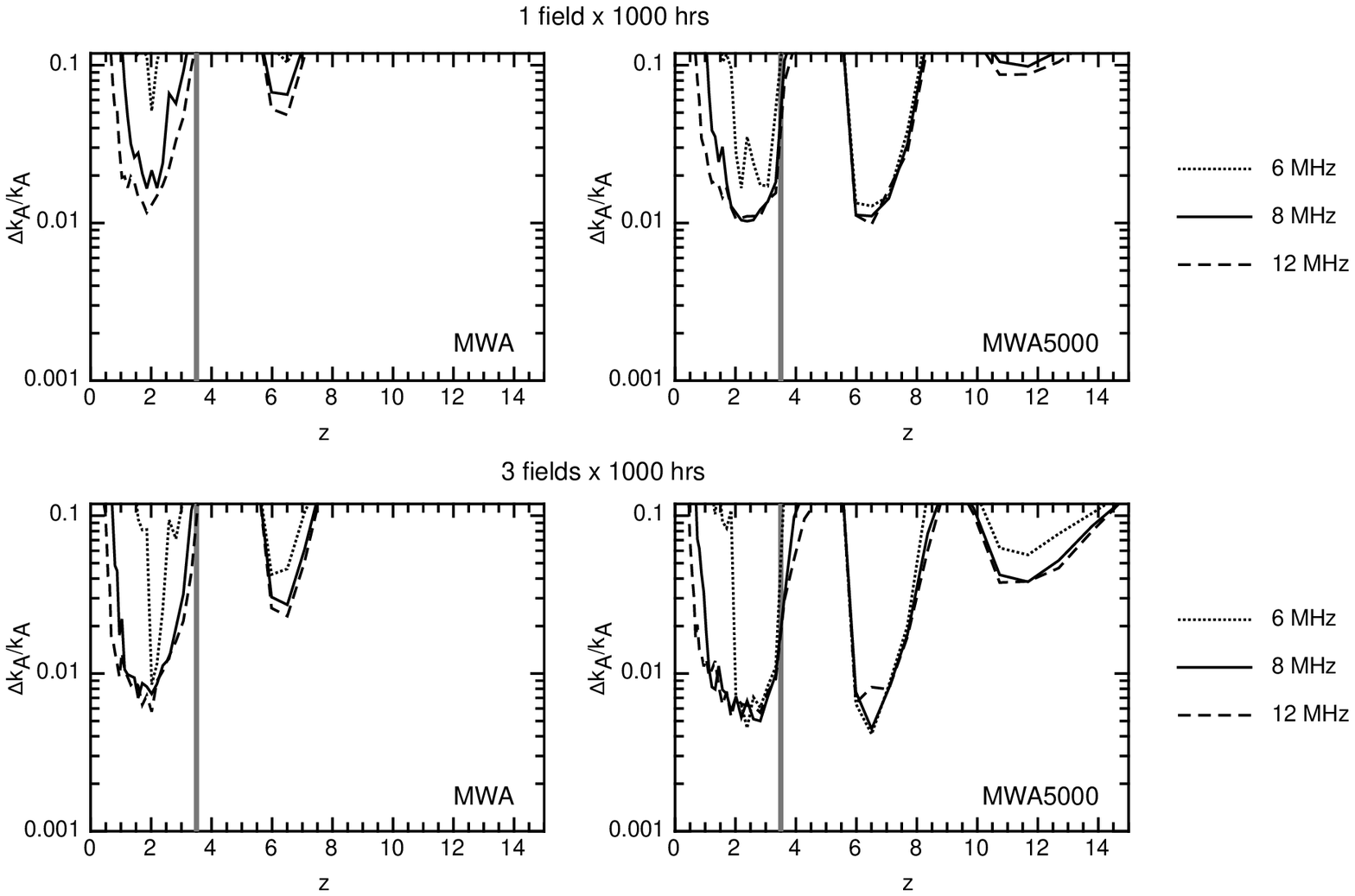} 
\caption{ The relative accuracy achievable on the measurement of $k_{\rm
A}$ as a function of redshift, assuming a prior probability on $k_{\rm A}$
of $k_{\rm A}>0.03$Mpc$^{-1}$. Results for the MWA (left panel) and the
MWA5000 (right panel), with foreground removal in $B=6$MHz (dotted lines),
$B=8$MHz (solid lines) and $B=12$MHz (dashed lines) band-passes
respectively.  The upper and lower panels show results assuming 1 and 3
fields, respectively (with each field observed for 1000hr). In each panel
we plot a vertical line at $z=3.5$. To the right of this line we assume
antennae with the specifications of the MWA. To the left of this line we
assume an antennae number equal to the MWA or MWA5000, but tiles with
$A_{\rm e}=A_{\rm tile}$.}
\label{fig5}
\end{figure*}

Figure~\ref{fig3} shows results for the power spectrum of 21cm
fluctuations, and sensitivity to the BAO scale at $z=3.5$, 5, 6.5 and
8. The spherically averaged model power spectra are marked on the left-hand
panels as the thick dark lines (these are only valid on large scales as
they do not capture the scale dependent bias that is the signature of the
HII bubbles on smaller scales). We also show (heavy grey lines) the
component of the power spectrum due to the baryonic acoustic oscillations.
(Note that we show the absolute value of the full power spectrum minus a
representative no wiggle power spectrum.) Estimates of the sample variance
(dotted lines) and thermal noise (dashed lines) components of the
uncertainty for detection by the Mileura Widefield Array (MWA) are plotted
in each of the panels. The MWA, which is currently under construction will
comprise a phased array of 500 tiles. Each tile will contain 16
cross-dipoles to yield an effective collecting area of $A_{\rm
e}=16(\lambda^2/4)$ (the area is capped for $\lambda>2.1$m). The tiles will
be distributed over an area with diameter 1.5km. The physical area of a
tile is $A_{\rm tile}=16$m$^2$. In this paper we consider 1000 tiles for
phase II of the MWA. We model the antennae distribution as having
$\rho(r)\propto r^{-2}$ with a maximum radius of 750m and a finite density
core of radius 20m, and we assume a 1000hr integration on a single field, a
band-pass over which foregrounds are removed of $B=8$MHz, and $k$-space
bins of width\footnote{The signal-to-noise is increased in proportion to
$\sqrt{\Delta k}$, and so will be substantially better per bin in
measurements of the power spectrum at lower resolution in $k$.}  $\Delta
k=k/10$. The total processed band pass for the MWA is $B_{\rm
total}=32$MHz. The combined uncertainty including the minimum $k$ cutoff
due to foreground subtraction is shown as the thin solid line.

The central panels of Figure~\ref{fig3} show the power spectra at $z=3.5$,
5, 6.5 and 8 with the representative smooth power spectrum subtracted. The
points with error bars show the accuracy attainable within a bin of width
$\Delta k/k=0.1$. The vertical dotted line is the wave number corresponding
to the band-pass, below which the error bars are very large. We have fitted
the analytic approximation\footnote{More recently a new technique has been
proposed (Percival et al.~2007; Angulo et al.~2007) which replaces the
analytic form of equation~(\ref{wanal}) with a scheme that uses a reference
power spectrum derived from the observed power spectrum, a full linear
perturbation theory power spectrum, plus a damping scale to account for
non-linear evolution. This method improves the fit to BAO power spectra
computed in numerical simulations, and provides a more general
approach. However we have chosen to employ the simpler approach of Blake \&
Glazebrook~(2003) in this initial investigation. } to the baryonic
oscillation component of the spherically averaged power spectrum following
Blake \& Glazebrook~(2003)
\begin{equation}
\label{wanal}
\frac{P(k)}{P_{\rm ref}(k)} = 1+ A k \exp{\left[-\left(\frac{k}{0.07\mbox{Mpc}^{-1}}\right)^{1.4}\right]}\sin\left(2\pi\frac{k}{k_{\rm A}}\right).
\end{equation}
This function has two parameters $A$ and $k_A$. The value of $A$ is
determined to high accuracy from observations of the CMB. For the purposes
of this analysis we therefore assume that $A$ is a known constant, and fit
only for $k_{\rm A}$. We fit only to values of $k<0.25$Mpc$^{-1}$. The
accuracy to which $k_{\rm A}$ can be measured determines the constraints
that 21cm power spectra can place on the dark energy. The right-hand panels of
Figure~\ref{fig3} show the probability distributions for the recovered
$k_{\rm A}$ at each redshift considered. In a single field the MWA could
detect the acoustic scale just prior to overlap, but could not make a
precise measurement (less than a few percent) at any redshift.

At values of $k\sim10^{-1}$ Mpc$^{-1}$, the measurement of
the power spectrum using the MWA will be limited by the thermal sensitivity
of the array, and so the signal-to-noise achievable in this regime will be
greatly enhanced by a subsequent generation of instruments with a larger
collecting area. As an example, we consider a hypothetical follow-up
instrument to the MWA which would comprise 5 times the total collecting
area.  We refer to this follow-up telescope as the MWA5000. The design
philosophy for the MWA5000 would be similar to the MWA, and we therefore
assume antennae distributed as $\rho(r)\propto r^{-2}$ with a diameter of
2km and a flat density core of radius 80m (see McQuinn et al.~2006).  In
Figure~\ref{fig4} we repeat our analysis of the power spectrum and baryonic
acoustic oscillations at $z=3.5$, 5, 6.5 and 8 for measurements using the
MWA5000. The panels show the same results as described in
Figure~\ref{fig3}. For the model overlapping at $z=6$, we find that the BAO
scale could be detected at at a range of redshifts and that very good
measurements of $k_{\rm A}$ ($\sim1\%$) could be made at $z=6.5$ with the
MWA5000 in a single field.

Following the results of McQuinn et al.~(2006) we do not make
estimates for the SKA\footnote{See www.skatelescope.org/} in this
paper. Current projections for the specifications of the SKA call for
large antennae, with a small fraction of collecting area concentrated
in a core. This design limits the field of view, as well as the
fraction of the telescope that can be used to measure the large scale
modes which probe the BAOs. Thus, despite its increased collecting
area, the SKA would be less powerful (with respect to measurement of
the redshifted 21cm power spectrum) than the MWA5000, whose design
would be optimized for the measurement of the 21cm power spectrum at
high redshift. We note that at some redshifts the MWA5000 would be
cosmic variance limited on scales relevant to BAO studies. As a
result if an SKA were built with a design based on the MWA5000, but
with ten times the collecting area, no substantial gains could be made
using observations of an individual field. Of course a telescope with
a larger collecting area could reach the limit of cosmic variance in a
shorter integration, allowing more fields to be observed.

To quantify the relative accuracy achievable on the measurement of $k_{\rm
A}$ we plot the variance of the recovered distribution divided by the best
fit value in Figure~\ref{fig5}.  In each panel of Figure~\ref{fig5} we plot
a vertical line at $z=3.5$. The MWA and MWA5000 could observe to the right
of this line, and we assume antennae with the specifications of the MWA in
this region. We note that there is non-zero probability for the recovered
$k_{\rm A}$ to lie at a harmonic of the true value (since the fitting
function is quasi-periodic and the data is noisy). This may be seen in the
distributions shown in Figures~\ref{fig3}-\ref{fig4}. For the results
presented in Figure~\ref{fig5} we assume a prior probability on $k_{\rm A}$
which is constant for $k_{\rm A}>0.03$Mpc$^{-1}$ but equal to zero
otherwise. This effectively assumes that we know the accuracy of $k_{\rm
A}$ to $\sim20\%$ apriori.  The upper left hand panel of Figure~\ref{fig5}
shows results for the MWA, with $B=6$MHz, $B=8$MHz and $B=12$MHz
band-passes respectively, and 1000hr of integration for a single field. If
foreground subtraction could be achieved, the larger band-passes would
improve the accuracy significantly at low $z$ by giving access to the peak
centered on $k\sim0.05$Mpc$^{-1}$. The MWA will not make a precision
measurement of the BAO scale using only one field. The upper right hand
panel of Figure~\ref{fig5} shows the corresponding results for MWA5000,
again for a single field and 1000hr of integration. In a single field the
MWA5000 could make precise measurements ($\Delta k_{\rm A}/k_{\rm
A}\sim1\%$) over an extended interval prior to overlap, but not at higher
redshifts.

Since any one observing field can only be observed for a fraction of the
time, measurement of the 21cm power spectrum will be performed over several
different fields. In addition, some phased arrays will have the capability
to observe using several primary beams at once. In the lower panels of
Figure~\ref{fig5} we show results that assume an integration time of 1000hr
on each of 3 separate fields (note that this corresponds to $\sim3$hr per
day per field for 1 year). The noise on the power spectrum scales as the
inverse square-root of the number of fields (McQuinn et al.~2006), which
results in an improved precision on measurements of $k_{\rm A}$ over that
achievable in a single field [see equations~(\ref{delsym}) and
(\ref{delasym})]. The precision achieved at $z\sim6.5$ would be as low as
$\Delta k_{\rm A}/k_{\rm A}\sim3\%$ using the MWA (comparable with the best
current measurements from galaxy surveys), while the MWA5000 could reach
$\Delta k_{\rm A}/k_{\rm A}\sim0.5\%$. In addition the MWA5000 could
precisely ($\Delta k_{\rm A}/k_{\rm A}\la 2\%$) measure the acoustic scale
at $z\sim3.5$.

The antenna design of the MWA is optimized for the epoch of reionization
measurements and only allows the 21cm line to be observed at
$z\ga3.5$. However, in Figure~\ref{fig5} we show results down to a redshift
of $z=0$ because a future instrument could be constructed with antennae
that are sensitive to a different frequency range. At $z\la3.5$ (to the
left of the vertical grey lines in Figure~\ref{fig5}) we assume $A_{\rm
e}=A_{\rm tile}$ and that the system temperature is dominated by the sky
(an assumption implying that, unlike the MWA, such a telescope will need to
have cooled receivers due to the lower sky temperature at shorter
wavelengths). An instrument constructed with the same number of antennae as
the MWA but which operated at a higher frequency range could accurately
measure the scale of acoustic oscillations at lower redshifts provided that
foreground subtraction could be achieved over a sufficiently large
band-pass. In constructing Figure~\ref{fig5} we have computed the 21cm
power spectrum assuming a constant value of $b_{21}$ at $z<2$. Our modeling
of $b_{21}$ does not include quasars and assumes local absorption of
ionizing photons. It therefore becomes unreliable at $z\la2$. However the
assumption of constant $b_{21}$ is reasonable given that the observed
density parameter of neutral gas does not vary significantly with redshift
(Prochaska et al.~2005). At low redshifts the precision for measurements of
$k_{\rm A}$ would be limited by the ability to remove foregrounds. The
minimum value of $k$ which is probed by the 21cm power spectrum becomes
larger as the observing frequency is increased, and at $z\la1$ has moved to
a value beyond the scale of the BAO peaks. Nevertheless, at $z\sim1.5$
observations from 3 fields could yield precisions of $\sim2\%$ and
$\sim1\%$ on measurements of $k_{\rm A}$ using 1000 and 5000 antennae
($A_{\rm e}=A_{\rm tile}$) respectively, with even higher precision
($\sim0.7\%$ and $\sim0.5\%$ respectively) attainable at $z\sim2.5$.

\subsection{Sensitivity to the transverse and line-of-sight acoustic scales}

\begin{figure*}
\vspace{10mm}
\includegraphics[width=15cm]{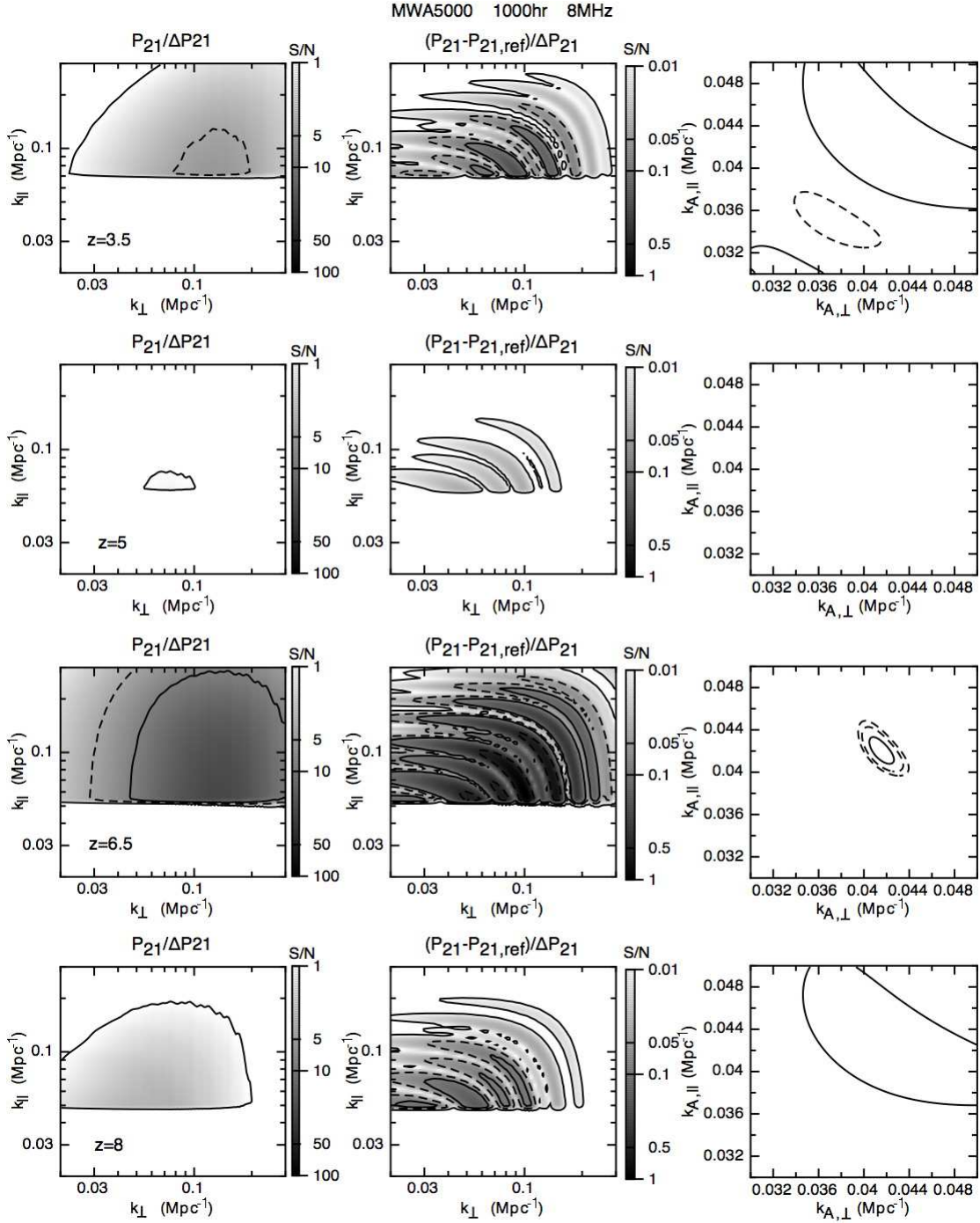} 
\caption{{\em Left panels:} The signal-to-noise ratio ($S/N$) for
observations of the power spectrum $P_{21}(k_{\perp},k_{\parallel})$. {\em
Central panels:} The signal-to-noise ratio ($S/N$) for observations of the
difference between the full power spectrum $P_{21}(k_{\perp},k_{\parallel})$
and the no wiggle reference power spectrum $P_{\rm
ref,21}(k_{\perp},k_{\parallel})$. In each case the signal-to-noise ratio has
been computed in bins of volume $2\pi k_{\perp} \Delta k_{\perp} \Delta
k_{\parallel}$, where $\Delta k_{\rm perp}=k_{\perp}/10$ and $\Delta
k_{\parallel}=k_{\parallel}/10$. {\em Right panels:} Contours of constant
likelihood for the recovered $k_{\rm A,\perp}$ and $k_{\rm A,\parallel}$
around the true input value. Contours are shown at values of
$\chi^2-\chi^2_{\rm min}=1$, 2.71 and 4, where $\chi^2_{\rm min}$ is the
value corresponding to the best fit parameter set. When projected onto
individual parameter axes ($k_{\rm A,\perp}$ and $k_{\rm A,\parallel}$),
the extrema of these contours represent the 68\%, 90\% and 95\% confidence
intervals on values of the individual parameters. The four rows show
results at $z=3.5$, $z=5$, $z=6.5$ and $z=8$.  The observational parameters
assume the layout and collecting area of the MWA5000 with 1000hr of
integration on a single field, a band-pass of 8MHz and redshifts of
$z=3.5$, $z=5$, $z=6.5$ and $z=8$.  }
\label{fig6}
\end{figure*}

In the previous section we have computed the sensitivity of 21cm
experiments to the angle-averaged value of $k_{\rm A}$. However,
observations of the 3-dimensional power spectrum of 21cm fluctuations
provide constraints on both the radial and transverse measures of this
scale. In Figure~\ref{fig6} we present results for the sensitivity of 21cm
observations to the line-of-sight and transverse BAO scale at $z=3.5$,
$z=5$, $z=6.5$ and $z=8$. In the left hand panels we show the
signal-to-noise for observations of the full power spectrum
$P_{21}(k_{\perp},k_{\parallel})$. In the central panels we show the
corresponding signal-to-noise for observations of the difference between
the full power spectrum $P_{21}(k_{\perp},k_{\parallel})$ and the no wiggle
reference power spectrum $P_{\rm 21,ref}(k_{\perp},k_{\parallel})$. For
this calculation we assume observation of a single field using the MWA5000,
1000hr of integration, and foreground removal within $B=8$MHz of
band-pass. The signal-to-noise has been computed in bins of volume $2\pi
k_{\perp} \Delta k_{\perp} \Delta k_{\parallel}$, where $\Delta
k_{\perp}=k_{\perp}/10$ and $\Delta k_{\parallel}=k_{\parallel}/10$. We
have used the analytical approximation from Glazebrook \& Blake~(2005)
\begin{eqnarray}
\nonumber
\frac{P(k)}{P_{\rm ref}(k)} &=& 1+ A k \exp{\left[-\left(\frac{k}{0.07\mbox{Mpc}^{-1}}\right)^{1.4}\right]}\\
&&\hspace{2mm}\times\sin\left[2\pi\sqrt{\left(\frac{k_{\perp}}{k_{\rm A,\perp}}\right)^2+\left(\frac{k_{\parallel}}{k_{\rm A,\parallel}}\right)^2}\right],
\end{eqnarray}
where $k^2=k_{\perp}^2+k_{\parallel}^2$, to estimate the corresponding
constraints on the line-of-sight and transverse acoustic scales ($k_{\rm
A,\parallel}$ and $k_{\rm A,\perp}$). On the right hand panels of
Figure~\ref{fig6} we show contours of likelihood for the recovered values
of $k_{\rm A,\perp}$ and $k_{\rm A,\parallel}$ around the true input value
for our standard cosmology. Figure~\ref{fig6} shows that redshifted 21cm
observations would be sensitive to both the transverse and line-of-sight
components of the oscillation scale, with comparable uncertainty.

\subsection{Comparison with galaxy surveys}

Constraints on the BAO acoustic scale may be used to constrain
parameters in models of dark energy, with the ability of a survey to
discriminate among different models of dark energy governed by the
accuracy achieved in measurements of the line-of-sight and transverse
acoustic scales. Before proceeding to discuss dark energy constraints,
we therefore first pause to compare the accuracy of the 21cm
experiment with potential galaxy redshift surveys. Glazebrook \&
Blake~(2005) present the simulated precision on measurements of
$k_{\rm A,\perp}$ and $k_{\rm A,\parallel}$ from hypothetical galaxy redshift
surveys. Spectroscopic surveys of $\sim10^6$ galaxies within $\Delta
z=0.5$ redshift bins in the range $1<z<3.5$ covering 1000 square
degrees would each measure the transverse and line-of-sight acoustic
scales to accuracies of $\sim1\%$ and $\sim2\%$ respectively. On the
other hand a photometric redshift survey over 2000 square degrees
between $2.5<z<3.5$ would measure the transverse scale to $\sim1\%$,
but would not constrain the line-of-sight scale. Similar results were
obtained at $1.5\la z \la 2.5$.

The SKA could also be used to do a galaxy survey of sufficient size to
measure the BAO scale (Abdalla \& Rawlings~2004), so long as it were
designed to have a sufficiently large field of view. However, the full
sensitivity of the SKA would be required to push the galaxy survey beyond
$z\sim1.5$, limiting the studies of BAO using galaxies to relatively low
redshifts when compared with optical spectroscopic surveys. By looking at
fluctuations in the surface brightness of unresolved 21cm emission, an
array like the MWA could push measurement of the BAO to much higher
redshift.

The very large areas of sky that must be surveyed in order to measure the
BAO scale using galaxy redshift surveys arise because very large volumes
must be sampled in order to beat down the statistical noise on the large
scale modes relevant to BAOs. For example, in units of the SDSS survey
volume ($\mathcal{V}_{\rm sloan}\sim5.8\times10^8$Mpc$^3$),
$\mathcal{V}/\mathcal{V}_{\rm sloan}\sim3$ and
$\mathcal{V}/\mathcal{V}_{\rm sloan}\sim1.8$ are required to achieve $2\%$
accuracy on $k_{\rm A}$ at $z\sim1$ and $z\sim3$ respectively (Blake \&
Glazebrook~2003). By comparison, in a single pointing within $B_{\rm
total}=32$MHz of band pass the MWA surveys $\mathcal{V}/\mathcal{V}_{\rm
sloan}\sim2.4$, $\mathcal{V}/\mathcal{V}_{\rm sloan}\sim4.8$,
$\mathcal{V}/\mathcal{V}_{\rm sloan}\sim7.6$ and
$\mathcal{V}/\mathcal{V}_{\rm sloan}\sim11$ at $z=1.5$, $z=2.5$, $z=3.5$
and $z=6.5$ (where we have assumed $A_{\rm e}=A_{\rm tile}$)
respectively. These volumes must be multiplied by $N_{\rm fields}$ in order
to get the total volume from which the power spectrum is to be constructed.
Thus observations of the 21cm power spectrum will probe very large volumes
of the IGM, comparable to the most ambitious galaxy redshift surveys.

We find that observations using a low frequency array the size of the MWA,
but with an appropriate frequency range would achieve measurements of the
acoustic scale at $z\sim1.5$--$2.5$ that are comparable in precision
($\sim1\%$) to high redshift galaxy surveys using the next generation of
optical instruments.  Moreover an instrument with the collecting area of
the MWA5000 would extend this sensitivity out to $z\sim3.5$.  Observations
of 21cm fluctuations would also allow the acoustic scale to be measured
with comparable precision at much earlier cosmic epochs ($z\sim6$), using
the MWA5000. Moreover, 21cm observations will be comparably sensitive to
the line-of-sight and transverse acoustic scales. In contrast,
spectroscopic galaxy redshift surveys are more sensitive to the transverse
scale, while photometric redshift surveys will not be sensitive at all to
the radial BAO (Glazebrook \& Blake~2005).

\section{Constraints on dark energy}
\label{DarkEnergy}

\begin{figure*}
\includegraphics[width=14cm]{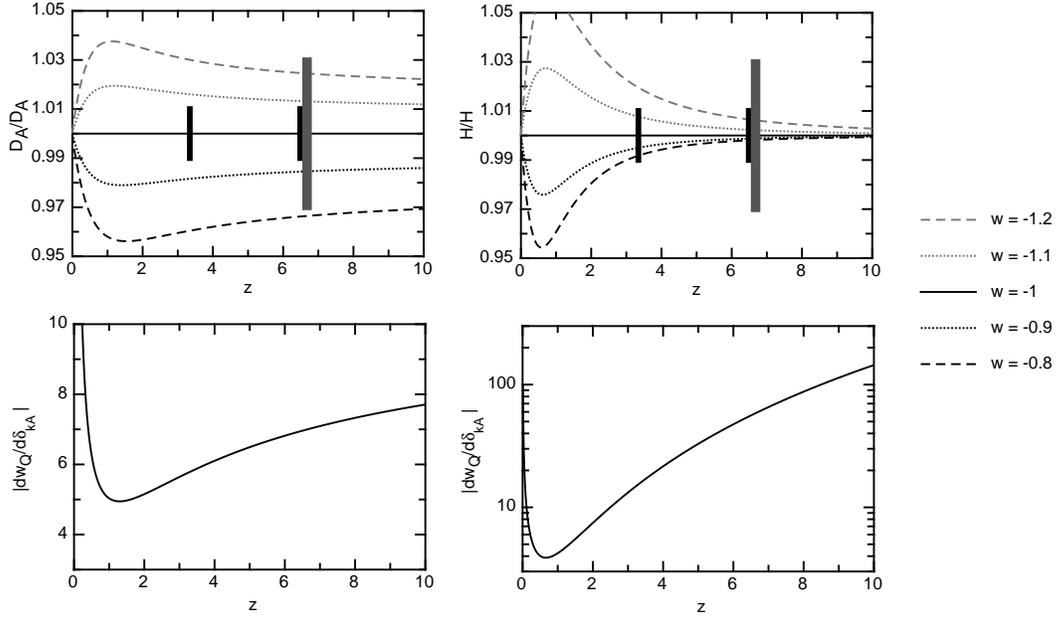} 
\caption{Constraints on a constant $w_Q$. {\em Upper panels:} The relative
change in angular diameter distance (left panel) and Hubble parameter
(right panel) as a function of redshift for different models of the
evolution in $w_Q$.  The fractional changes are relative to a cosmological
constant ($w_Q=-1$).  The thick black and grey bars illustrate an accuracy
of 1\% and 3\% on $k_{\rm A}$ which would be achievable in some redshift
ranges in a 1000hr observation by the MWA5000 (1
field) and MWA (3 fields) respectively. {\em Lower panels:} The uncertainty
in $w_Q$ per unit relative uncertainty in the angular diameter distance
$D_{\rm A}$ (left panel) and Hubble parameter $H$ (right panel), as a
function of redshift. The uncertainty is relative to a cosmological
constant ($w_Q=-1$).}
\label{fig7}
\end{figure*} 

The measurement of the angular scale of acoustic oscillations
(i.e. oscillations transverse to the line-of-sight) provides a measure of
the angular diameter distance at a redshift $z$ where the oscillations are
observed. The angular diameter distance can also be computed at a redshift
$z$ for a chosen cosmology
\begin{equation}
D_{\rm A} = (1+z)^{-1}\int_0^z {dz^\prime\over H(z^\prime)},
\label{DA}
\end{equation}
where $H(z)$ is the Hubble parameter at redshift $z$, and we have set the
speed of light equal to unity. Thus the measurement of $D_{\rm A}(z)$
probes the underlying geometry of the universe. On the other hand,
measurement of the redshift scale of acoustic oscillations
(i.e. oscillations along the line-of-sight) provides a direct measure of the
Hubble parameter at a redshift $z$, which may be written
\begin{eqnarray}
\nonumber
H(z) &=& -H_0(1+z)^{3/2}\\
&&\hspace{-5mm}\times\left[\Omega_m +
\Omega_{Q}\exp{\left(3\int_0^z\frac{dz^\prime}{(1+z^\prime)}w_Q(z^\prime)\right)}\right]^{1/2}.
\end{eqnarray}
In writing the Hubble parameter we have parameterized the equation of state as,
\begin{equation}
p_i = w_i(z)\rho_i,
\end{equation}
where the subscript $i$ refers to either pressureless matter
(with $w=0$ and a subscript $i=m$) or dark energy (with a subscript $i=Q$).  
We have also used the energy conservation for the dark energy
$\dot{\rho}_Q=-3H(z)[1+w_Q(z)]\rho_Q$. 
The redshift dependence of the acoustic scale has been suggested as a
powerful probe of the evolution of the dark energy (Blake \&
Glazebrook~2003; Hu \& Haiman~2003; Seo \& Eisenstein~2003; Glazebrook \&
Blake~2005; Seo \& Eisenstein 2007; Angulo et al. 2007). If the size of the
sound horizon is known, then the relative errors $\Delta D_{\rm A}/D_{\rm
A}$ in the angular diameter distance and $\Delta H/H$ in the Hubble
parameter are related to the relative errors $\Delta k_{\rm A,\perp}/k_{\rm
A,\perp}$ and $\Delta k_{\rm A,\parallel}/k_{\rm A,\parallel}$ in the
transverse and line-of-sight acoustic scales respectively
(Figure~\ref{fig6}).

To investigate the utility of the redshifted 21cm emission as a probe
of the dark energy equation of state we adopt a simple approach. In
the remainder of this section we assume that the values of $\Omega_m$,
$\Omega_Q$ and the present-day value of the Hubble parameter $H_0$
have been precisely determined apriori, so that we do not consider
the uncertainty in the sound horizon. We also assume that the Universe
is flat. In the upper panels of Figure~\ref{fig7} we plot the
fractional change in the angular diameter distance and Hubble
parameter respectively for several models with constant $w_Q$,
relative to a model with a cosmological constant ($w_Q=-1$).  The
vertical bars illustrate the type of precision achievable by the MWA
(grey bar) and MWA5000 (dark bars) in a 1000hr observation of three
fields and a single field respectively, at the redshifts where these
observations will be most efficient. Based on the transverse
measurement of the BAO scale, the MWA would be able to make a
measurement of $w_Q$ with $\sim20\%$ accuracy using a single
measurement at $z\sim6.5$, while the MWA5000 could make a measurement
of $w_Q$ at the $\sim7\%$ level from observations at each of
$z\sim3.5$ and $z\sim6.5$, yielding a combined constraint of better
than $5\%$. Better precision could be achieved using observations of three
fields. Since a cosmological constant does not affect the dynamics at
high redshift the line-of-sight BAO scale is left unaffected by this
change in the equation of state. Thus a cosmological constant would
introduce an asymmetry in the BAO scale that would be easily detected
by high redshift observations.

We can obtain a more quantitative relation between the uncertainties in
$k_{\rm A}$ and $w_Q$. If we expand around $w_Q=-1$, then in the case of a
constant $w_Q$ model we can relate the uncertainty in $w_Q$ to the observed
uncertainty in the scale $k_{\rm A}$. We obtain
\begin{equation}
\Delta w_Q = \left| \frac{dw_Q}{d\delta_{kA}} \right|\delta_{kA}^{\rm obs},
\end{equation}
where $\delta_{kA}^{\rm obs}\equiv\Delta k_{\rm A,\perp}/k_{\rm A,\perp}$
or $\delta_{kA}^{\rm obs}\equiv\Delta k_{\rm A,\parallel}/k_{\rm
A,\parallel}$ for the transverse and line-of-sight components
respectively. In the lower left panel of Figure~\ref{fig7} we plot the
resulting uncertainty in $w_Q$ per unit fractional uncertainty in the
transverse measurement of $k_{\rm A}$. This figure shows that a 10\%
uncertainty in $w_Q$ requires $k_{\rm A}$ to be measured in the transverse
direction with a precision of $\sim1.5\%$. As mentioned above this will be
achievable at each of redshifts $z\sim3.5$ and $z\sim6$ for the
MWA5000. The small change in the radial BAO scale with $w_{Q}$ is reflected
by the large value of the derivative in the lower right panel of
Figure~\ref{fig7}.

\begin{figure*}
\includegraphics[width=13cm]{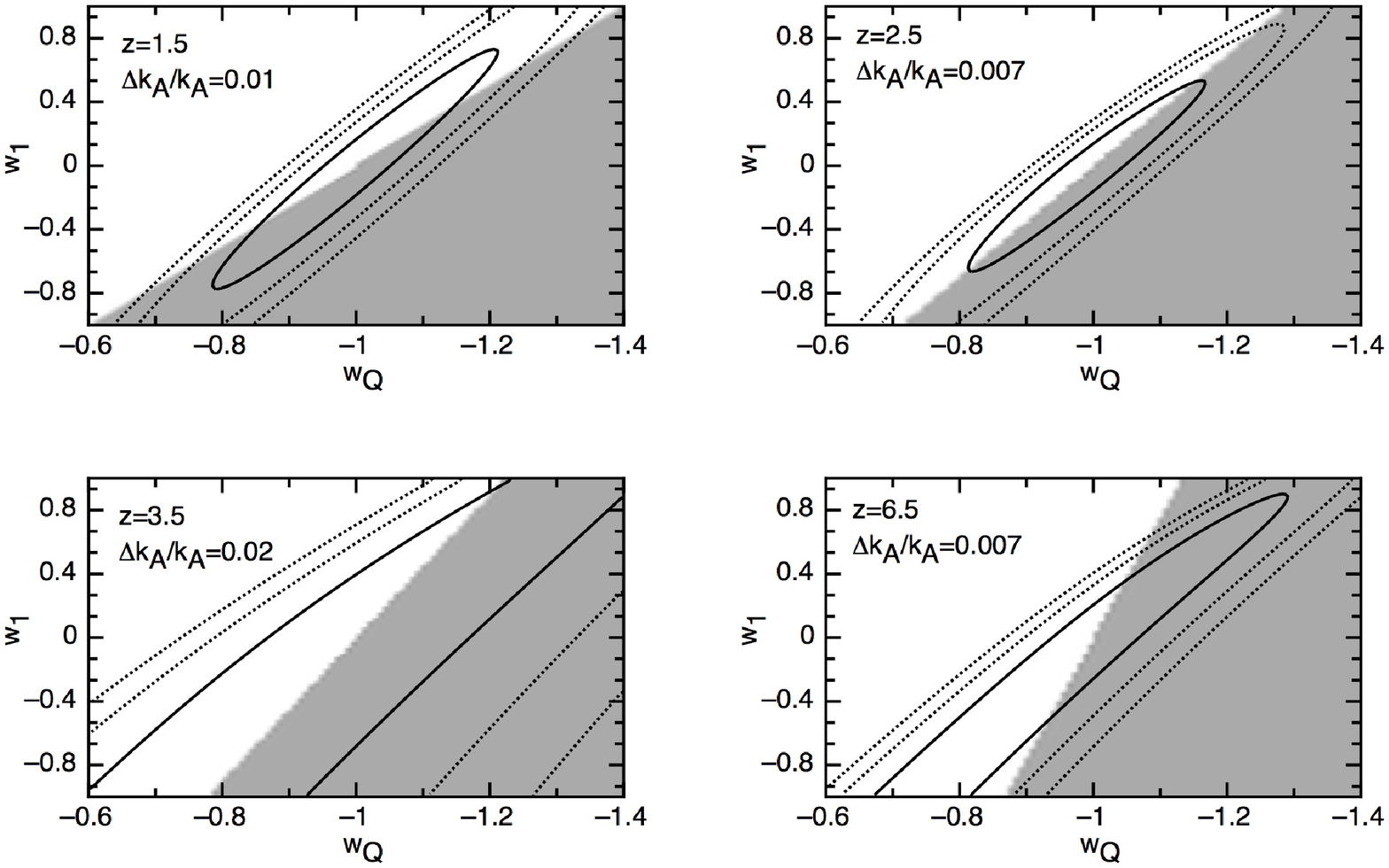} 
\caption{The joint likelihoods for $w_Q$ and $w_1$ relative to a model with
a cosmological constant ($w_Q=-1$, $w_1=0$). The contours are at 64\%, 26\%
and 10\% of the maximum and were found from combining constraints on
$D_{\rm A}$ and $H$, assuming uncertainties $\Delta k_{\rm A,\perp}/k_{\rm
A,\perp}=\Delta k_{\rm A,\parallel}/k_{\rm A,\parallel}=\sqrt{2}\times0.01$
(at $z=1.5$, upper left panel), $\Delta k_{\rm A,\perp}/k_{\rm
A,\perp}=\Delta k_{\rm A,\parallel}/k_{\rm
A,\parallel}=\sqrt{2}\times0.007$ (at $z=2.5$, upper right panel), $\Delta
k_{\rm A,\perp}/k_{\rm A,\perp}=\Delta k_{\rm A,\parallel}/k_{\rm
A,\parallel}=\sqrt{2}\times0.02$ (at $z=3.5$, lower left panel), and
$\Delta k_{\rm A,\perp}/k_{\rm A,\perp}=\Delta k_{\rm A,\parallel}/k_{\rm
A,\parallel}=\sqrt{2}\times0.007$ (at $z=6.5$, lower right panel). At
$z\geq3.5$ these sensitivities correspond to three fields, each observed
for 1000hr with the MWA5000. At $z\leq2.5$ these sensitivities correspond
to three fields, each observed for 1000hr with an array having 5000
antennae (each with 16 cross-dipoles) and $A_{\rm e}=A_{\rm tile}$. Models
with $w_Q(z)<-1$ or $w_Q(z)>1$ violate the weak energy condition, and we
shade these regions grey.}
\label{fig8}
\end{figure*} 

\subsection{Constraints on parameterized models of evolving dark energy}

In the remainder of this section we constrain two different parameterized
dark energy models assuming observations with the MWA5000 at four different
redshifts.  Rather than consider the spherically averaged constraint on
$k_{\rm A}$, in this section we utilize independent constraints of the
line-of-sight and transverse BAO scales (Glazebrook \&
Blake~2005). Figure~\ref{fig6} illustrates that the MWA5000 will have a
comparable sensitivity to measurements of $k_A$ along the line-of-sight
($k_{\rm A,\parallel}$) as compared to perpendicular to it ($k_{\rm
A,\perp}$). We therefore assume the transverse and line-of-sight
sensitivities to be $\Delta k_{\rm A,\perp}/k_{\rm A,\perp}=\Delta k_{\rm
A,\parallel}/k_{\rm A,\parallel}=\sqrt{2}\times \Delta k_{\rm A}/k_{\rm
A}$, and constrain dark energy parameters using the joint constraints on
$D_{\rm A}$ and $H$ (i.e. we assume $\Delta k_{\rm A,\perp}/k_{\rm
A,\perp}=\Delta D_{\rm A}/D_{\rm A}$ and $\Delta k_{\rm A,\parallel}/k_{\rm
A,\parallel}=\Delta H/H$). Note that our relations between the transverse
or line-of-sight sensitivities and the spherically averaged sensitivity to
the BAO scale assume that the transverse and line-of-sight scales are
independent, whereas Figure~\ref{fig6} illustrates that there is some
degeneracy. We therefore slightly underestimate the uncertainties on the
transverse and line-of-sight BAO scales.

The equation of state of dark energy may not be constant. The possible
evolution is often parameterized using the following simple form
\begin{equation}
\label{weqn}
w_Q(z)=w_Q+w_1(1-a),
\end{equation}
where $w_1$ is the derivative of $w_Q(z)$ with respect to the scale-factor
$a$.  We have investigated the joint constraints that the 21cm BAO
observations might place on the parameters $w_Q$ and $w_1$, and present the
results in Figure~\ref{fig8}. We construct the relative likelihoods for
$w_Q$ and $w_1$ given a true model with $w_Q=-1$ and $w_1=0$,
\begin{equation}
\mathcal{L}(w_Q,w_1)=\exp\left\{-\frac{1}{2}\left[\left(\frac{\delta k_{\rm
A,\perp}}{\delta k^{\rm obs}_{\rm A,\perp}}\right)^2+\left(\frac{\delta k_{\rm A,\parallel}}{\delta k^{\rm obs}_{\rm
A,\parallel}}\right)^2 \right]\right\},
\end{equation}
where $\delta k_{\rm A,\perp}(z)$ and $\delta k_{\rm A,\parallel}(z)$ are
functions of $w_Q$ and $w_1$. We show results in 4 redshift bins and have
assumed $\delta k^{\rm obs}_{\rm A,\perp}=\delta k^{\rm obs}_{\rm
A,\parallel}=\sqrt{2}\delta k^{\rm obs}_{\rm A}$, with observed
uncertainties of $\delta k^{\rm obs}_{\rm A}\equiv \Delta k_{\rm A}/k_{\rm
A}=0.01$ at $z=1.5$, $\delta k^{\rm obs}_{\rm A}=0.007$ at $z=2.5$, $\delta
k^{\rm obs}_{\rm A}=0.02$ at $z=3.5$, and $\delta k^{\rm obs}_{\rm
A}=0.007$ at $z=6.5$. The latter two cases are examples of the levels of
precision in $\Delta k_{\rm A}/k_{\rm A}$ that may be achievable in 1000hr 
integrations over each of three fields with the MWA5000, while the cases
at $z\sim1.5$ and $z\sim2.5$ correspond to an instrument with the antenna
number of the MWA5000, but with a higher frequency range and $A_{\rm
e}=A_{\rm tile}$ (again over three fields and 1000 hours). In each case
contours of the likelihood are shown at 64\%, 26\% and 10\% of the
maximum. The contours illustrate the degeneracy in $w_0$ and $w_1$ from a
single measurement of the acoustic scale at high redshift. This degeneracy
arises if the cosmological constant is the true model, because the effect
of dark energy is limited to the value observed for $D_{\rm A}$. However
both $w_Q$ and $w_1$ could be constrained at $z\la2.5$.  We note that
models with $w_Q(z)<-1$ or $w_Q(z)>1$ violate the weak energy condition,
and shade these regions grey in Figure~\ref{fig8}.  However, we include
constraints on all parameters for generality.

Constraints on the parameterization of the dark energy in
equation~(\ref{weqn}) are most accessible to observations at
$z\la2.5$. Indeed by construction equation~(\ref{weqn}) does not lead to
differences in the expansion history at $z\ga3.5$ where the BAO scale is
accessible to the MWA5000. Nevertheless, the combination of observations at
several redshifts at $z\ga3.5$ would provide tight limits on the value of
either $w_Q$ ($\la5\%$) if it is constant, or $w_1$ ($\pm0.2$) if $w_Q$ is
known apriori. In constructing these limits we have utilized external
information through our assumption that the cosmological parameters
$\Omega_m$, $\Omega_b$ and $\Omega_Q$ are known, but have not incorporated
additional sources of constraint on the dark energy equation of state at
lower redshift, which could include observations of Type~Ia supernovae
(e.g. Riess et al. 2007; Zhao et al. 2007), or the BAO scale from galaxy
redshift surveys (Cole et al.~2005; Eisenstein et al.~2005). The
constraints on the dark energy equation of state presented in
Figure~\ref{fig5} could of course be improved through a joint analysis of
all available constraints. For example, while the BAO scale measured by
Eisenstein et al.~(2005) does not by itself constrain the parameters of
dark energy, Wood-Vasey et al.~(2007) have combined this measurement with
observations of Type~Ia supernovae from the ESSENCE supernova survey, and
find (assuming a flat Universe) that $w_Q=-1.05^{+0.13}_{-0.12}$ with a
systematic uncertainty of $\pm0.11$.

\begin{figure*}
\includegraphics[width=13cm]{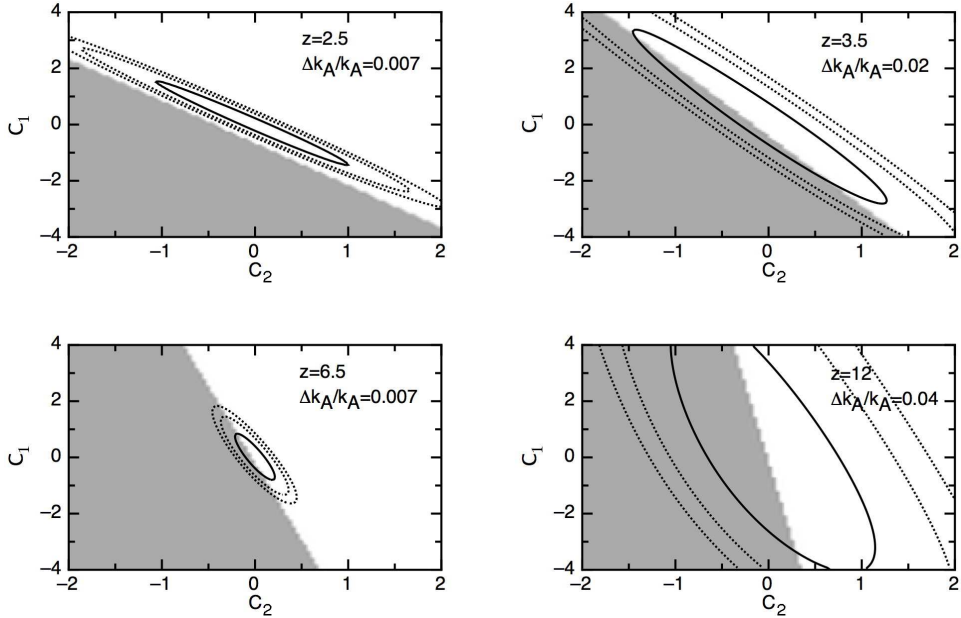} 
\caption{The joint likelihoods for $C_1$ and $C_2$ relative to a model with
a cosmological constant ($C_1=C_2=0$). The contours are at 64\%, 26\% and
10\% of the maximum and were found from combining constraints on $D_{\rm
A}$ and $H$, assuming uncertainties $\Delta k_{\rm A,\perp}/k_{\rm
A,\perp}=\Delta k_{\rm A,\parallel}/k_{\rm
A,\parallel}=\sqrt{2}\times0.007$ (at $z=2.5$, upper left panel), $\Delta
k_{\rm A,\perp}/k_{\rm A,\perp}=\Delta k_{\rm A,\parallel}/k_{\rm
A,\parallel}=\sqrt{2}\times0.02$ (at $z=3.5$, upper right panel), $\Delta
k_{\rm A,\perp}/k_{\rm A,\perp}=\Delta k_{\rm A,\parallel}/k_{\rm
A,\parallel}=\sqrt{2}\times0.007$ (at $z=6.5$, lower left panel), and
$\Delta k_{\rm A,\perp}/k_{\rm A,\perp}=\Delta k_{\rm A,\parallel}/k_{\rm
A,\parallel}=\sqrt{2}\times0.04$ (at $z=12$, lower right panel). At
$z\geq3.5$ these sensitivities correspond to three fields, each observed for
1000hr with the MWA5000. At $z=2.5$ these sensitivities correspond to three
fields, each observed for 1000hr with an array having 5000 antennae (each with
16 cross-dipoles) and $A_{\rm e}=A_{\rm tile}$. Models with $\rho_v<0$ are shaded grey.}
\label{fig9}
\end{figure*} 

Equation~(\ref{weqn}) imposes a certain type of evolution on the dark
energy. In particular, for all values of $w_1$ the parameterization in
equation~(\ref{weqn}) assumes that 50\% of the evolution in $w_Q(z)$
occurred at $z<1$. Moreover, the constraints on model parameters are only
meaningful should the model be a correct description of reality. Since 21cm
observations of BAO will be a powerful probe of dark energy at
high redshifts, we require a parameterization that is more general than
is provided by equation~(\ref{weqn}).  Current observations indicate that
the evolution of the cosmic expansion is consistent to within $\sim 10\%$
with a pure cosmological constant at $z\la1$ (Riess et al. 2007; Zhao et
al. 2007), while redshifted 21cm observations might be the first working
method of probing the dark energy at $z\ga3$. An alternative
parameterization to equation~(\ref{weqn}) for the possible evolution of the
vacuum energy density [$\rho_v(z)$] at high redshifts is to adopt a
constant value ($\rho_{v,0}$) in the redshift range $0\leq z\leq z_{\rm
min}$ which is constrained by other observational probes (such as Type Ia
supernovae or galaxy redshift surveys), and to represent the unknown
evolution at higher redshifts $z>z_{\min}$ by
\begin{equation}
\rho_v = \rho_{v,0}\left[ 1 + C_1(z-z_{\rm min}) + 
C_2(z-z_{\rm min})^2 + ...\right],
\label{rho_v}
\end{equation}
where we have kept only the two leading terms in the polynomial
expansion of $\rho_v(z)$. We next require that the Universe be flat, which
yields the corresponding evolution for the Hubble parameter at $z>z_{\rm
min}$,
\begin{eqnarray}
\nonumber
\left[{H(z)\over H_0}\right]^2&=&\Omega_{m,0}(1+z)^3+\Omega_{v,0}
[1+ C_1(z-z_{\rm min})\\
&&\hspace{23mm}+C_2(z-z_{\rm min})^2 + ...],
\label{Hz}
\end{eqnarray}
where the subscript $0$ denotes present-day values. This expression
 can then be substituted directly into
equation~(\ref{DA}) to obtain the angular diameter distance, $D_A(z)$. If
the vacuum energy density originates from a scalar field $\phi$ rolling
down a potential $V(\phi)$, then measurements of the above expansion
parameters can be used to infer the shape of $V(\phi)$ (see Linder 2007 and
Turner \& Huterer 2007 for recent reviews).

We have investigated the joint constraints that the 21cm BAO observations
might place on the parameters $C_1$ and $C_2$, and present the results in
Figure~\ref{fig9}. We do not constrain $\rho_{v}$ to be positive, but shade
the regions where $\rho_v<0$.  For this figure we assume $z_{\rm min}=1$
and that the dark energy at $z<z_{\rm min}$ is a cosmological constant. We
construct the relative likelihoods for $C_1$ and $C_2$ given a true model
with $C_1=C_2=0$.  As before we show results in four redshift bins and have
assumed observed uncertainties of $\delta k^{\rm obs}_{\rm A,\perp}=\delta
k^{\rm obs}_{\rm A,\parallel}=\sqrt{2}\delta k^{\rm obs}_{\rm A}$ with
$\delta k^{\rm obs}_A\equiv \Delta k_{\rm A}/k_{\rm A}=0.007$ at $z=2.5$,
$\delta k^{\rm obs}_A\equiv \Delta k_{\rm A}/k_{\rm A}=0.02$ at $z=3.5$,
$\Delta k_{\rm A}/k_{\rm A}=0.007$ at $z=6.5$, and $\Delta k_{\rm A}/k_{\rm
A}=0.04$ at $z=12$. In each case contours of the likelihood are shown at
64\%, 26\% and 10\% of the maximum. Since this model does not exclude a
dark energy contribution to the expansion dynamics at high redshift, its
parameters are well constrained by high redshift observations of the BAO
scale.  The contours illustrate the level of degeneracy in $C_1$ and $C_2$
from a single measurement of the acoustic scale at high redshift. The
combination of constraints at different redshifts would reduce this
degeneracy.

\section{Discussion}
\label{conclusion}

In this paper we have shown that measurements of the power spectrum
of fluctuations in the intensity of 21cm emission 
could provide a precise determination of the scale of the acoustic
peak in the matter power spectrum. By using the acoustic scale to
measure the angular diameter distance and Hubble parameter, it would
therefore be possible to use redshifted 21cm studies to constrain the
dark energy in the unexplored redshift range of $1.5\la z\la 15$.

The main advantage of 21cm observations is that in addition to studies
in the redshift range where a cosmological constant is expected to
dominate the expansion dynamics, measurement of the BAO scale would
also be possible at much earlier cosmic epochs than are accessible by
other methods.  Indeed, observations of the redshifted 21cm power
spectrum at $3.5\la z\la 12$ using a facility with ten times the
collecting area of the MWA would make a detailed study of the presence
of dark energy at high redshift that differs from that measured at
$z<1$.  Moreover, 21cm observations will be comparably sensitive to
the line-of-sight and transverse acoustic scales. In contrast,
spectroscopic galaxy redshift surveys are more sensitive to the
transverse scale, while photometric redshift surveys will not be
sensitive to the radial BAO at all (Glazebrook \& Blake~2005).

We note that a measurement of the power
spectrum both along the observer's line-of-sight as well as perpendicular
to it can also be used to constrain cosmological parameters through the
Alcock-Paczynski (1979) test in a 21cm data set (Barkana \& Loeb 2005a;
Nusser 2005; Barkana 2005). However, the use of the BAO scale as a cosmic
yardstick allows the BAO to be a more powerful probe of dark energy than
the Alcock-Paczynski~(1979) test. This is because the line-of-sight and
transverse measurements of the BAO scale yield independent measurements of
the expansion history, whereas the Alcock-Paczynski~(1979) test utilizes
only their ratio.

The redshift range accessible to an instrument like the MWA is
specified by the antennae design. A purpose built instrument with
sensitivity to a higher range of frequencies could probe the BAO scale
at lower redshifts, enabling continuous coverage from the reionization
era to $z\la1$ where galaxy and supernovae surveys become effective.
Redshifted 21cm studies of acoustic oscillations could in principle
also be extended to the redshift regime $15\la z\la 200$ in the more
distant future (Barkana \& Loeb 2005b). However, such an extension
would be particularly challenging because of the increase in the
Galactic synchrotron foreground towards lower observed frequencies.

We have shown that the first generation of low-frequency experiments
(such as MWA or LOFAR) will be able to constrain the scale of the
acoustic oscillations to within $\sim3\%$ in a high redshift window
just prior to overlap (see Fig. \ref{fig5}). This sensitivity to the
acoustic peak is comparable to the best current measurements from
galaxy surveys (Eisenstein et al.~2005). In addition, we find that
observations using an array 10 times the size of the MWA, and with
sensitivity in an appropriate frequency range would achieve
measurements of the BAO scale with precisions that are below one
percent in some particular redshift intervals. These windows include
the reionization era, as well as the range $1.5\la z\la 3.5$, where
the precision obtained would be comparable to ambitious high redshift
spectroscopic surveys of $\sim10^6$ galaxies, covering more than 1000
square degrees (Glazebrook \& Blake~2005).  Moreover, since
observations of the BAO scale in 21cm fluctuations can be extended
down to $z\sim1.5$ there will be an opportunity to demonstrate that the
results match those from other well studied techniques.  However, at
$z\ga3.5$, where the acoustic peak scale is not accessible through
standard techniques, the 21cm power spectrum would provide unique
constraints on the dark energy.

\bigskip

{\bf Acknowledgments} The authors thank Chris Blake, Matt McQuinn and Brian
Schmidt for comments which have greatly contributed to this work.  The
research was supported in part by the Australian Research Council (JSBW)
and Harvard University grants (AL). PMG acknowledges the support of an
Australian Postgraduate Award.

\newcommand{\noopsort}[1]{}

\label{lastpage}
\end{document}